\documentclass[12pt]{iopart}
\usepackage[dvips]{graphicx}
\usepackage{setstack, amsthm, amssymb, color, cite}
\usepackage{mathrsfs}

\begin{document}

\title[Anti-symmetric exclusion process]{An anti-symmetric exclusion process for two particles on an infinite 1D lattice}

\author{J R Potts$^{1,2}$, S Harris$^2$ and L Giuggioli$^{1,2,3}$}
\ead{jonathan.potts.08@bris.ac.uk}
\address{1. Bristol Centre for Complexity Sciences, University of Bristol, Bristol, UK. }
\address{2. School of Biological Sciences, University of Bristol, Bristol, UK. }
\address{3. Department of Engineering Mathematics, University of Bristol, Bristol, UK. }

\begin{abstract}
A system of two biased, mutually exclusive random walkers on an infinite 1D lattice is studied whereby the intrinsic bias of one particle is equal and opposite to that of the other.  The propogator for this system is solved exactly and expressions for the mean displacement and mean square displacement (MSD) are found.  Depending on the nature of the intrinsic bias, the system's behaviour displays two regimes, characterised by (i) the particles moving towards each other and (ii) away from each other, both qualitatively different from the case of no bias.  The continuous-space limit of the propogator is found and is shown to solve a Fokker-Planck equation for two biased, mutually exclusive Brownian particles with equal and opposite drift velocity.
\end{abstract}
\submitto{\JPA}
\maketitle

\section{Introduction and motivation}
\label{sec1}

Systems of randomly moving agents that exclude one another from the space they occupy are ubiquitous in science and technology, from RNA transcription \cite{RNA1, RNA2}, to territorial behaviour in the animal kingdom \cite{GPH1} to wireless networking \cite{wirelessnet}.  The theory of exclusion processes has been studied since 1965, when Harris \cite{exclusion1965} showed that a tagged particle, or tracer, on a 1D line subdiffuses at long times.  Since then, there have been a variety of mathematical developments of these so-called single-file systems \cite{EP1,EP2,EP3,EP4,EP5} whereby particle motion is overdamped and interaction is mutually exclusive.  

When the mutually excluding particles are unbiased, one often talks about systems undergoing symmetric exclusion \cite{SEP}.  On the other hand, if the particles are subject to a drift, one talks about asymmetric exclusion \cite{EP1}.  However, in all cases studied so far, symmetric or asymmetric, the particles undergoing exclusion exhibit identical behaviours.  In \cite{Aslangul} Aslangul solved exactly a particular symmetric exclusion process: the case of two unbiased repulsive random walkers on an infinite 1D lattice.  Here, we extend that work to the case where each walker does have an intrinsic bias, but the bias of one is {\textit{anti-symmetric}} to the other.  That is, the probability of the left-hand particle jumping right (left) at each step is $0<p<1$ ($1-p$) and the probability of the right-hand particle jumping left (right) is also $p$ ($1-p$).  

The practical motivation for our study arises from the collective emergence of territorial patterns in animal populations \cite{GPH1}.  Animals are called territorial if they each defend a region of space from possible intruders or neighbours.  Since they need to move around to carry out their vital activities such as foraging, animals are unable to monitor their territory boundaries on a permanent basis.  For this reason, many species have evolved an ability to define their territories using scent marking, thereby eschewing the need for continuous border patrolling.  An animal marks the terrain it visits by depositing a recognisable olfactory cue that is considered to be `active' by conspecifics for a finite amount of time.  As neighbours encounter active foreign scent, they move away to avoid costly confrontation.  

By modelling animals as {\it territorial random walkers} \cite{GPH2}, that is random walkers with such a scent-mediated interaction process, the terrain naturally subdivides into territories, demarcated by the area that contains active scent.  In 1D, each territory is a finite interval joined to adjacent territories at what we call the {\it borders}.  Since the scent is only active for a finite time, unless the animal re-scents its borders within this time, the borders will move.  Thus the borders can be viewed as randomly moving particles in their own right.  In addition, since smaller-than-average territories in the model end up having their borders re-scented more frequently than larger ones, they will tend to grow, whereas larger-than-average territories tend to shrink, meaning that the borders can be thought of as randomly moving particles connected by springs.  

\begin{figure}[htb]
\includegraphics[width=\columnwidth]{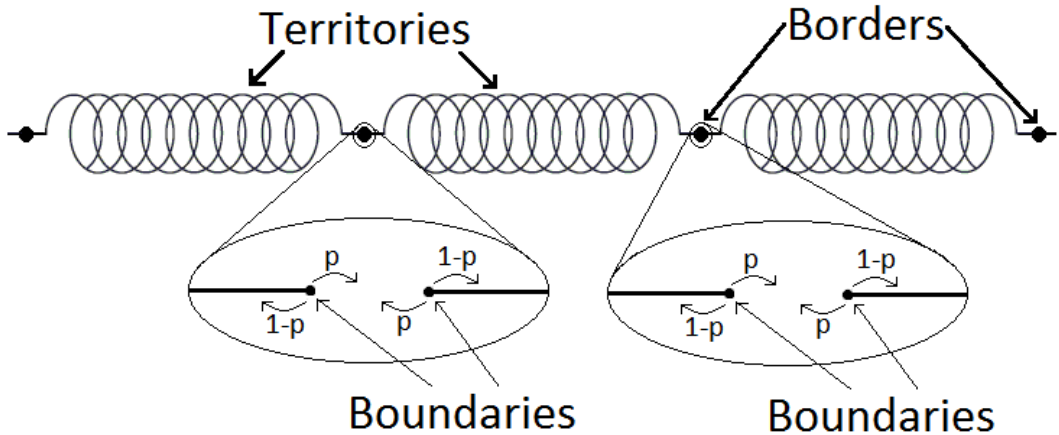}
\caption{Diagram of a model of territorial dynamics that reduces the interacting particle model of \cite{GPH1}.  Territories are modelled as springs, as in \cite{GPH2}, joined together by borders that are modelled as diffusive particles.  Zooming in on a border point reveals that it consists of two boundaries, each of which is moving randomly but with a drifting tendency towards the other.}
\label{territory_explanation}
\end{figure}

In figure \ref{territory_explanation}, we sketch a mathematical representation of the territories.  Each spring represents a territory, whose width fluctuates around a mean length equal to the inverse of the animal population density.  Each border is a particle whose movement is intrinsically random, though also constrained by the presence of the connected springs.  Consequently, since this is a form of symmetric exclusion process, the resultant movement of a tagged border particle is subdiffusive \cite{lizana2010}. 

However, by zooming in on a border one realises that it is actually made of two {\it boundaries}, one for each of the two adjacent territories.  The process by which the movement of these two boundaries gives rise to the intrinsic random movement of the border can be described by our present analysis of anti-symmetric random walkers and provides the main motivation for this work.  

The paper is organised as follows.  The model description and its exact solution form section \ref{sec2}.  In section \ref{sec3} long-time dependences are studied and compared with stochastic simulations, whereas the spatial continuum limit is analysed in section \ref{sec4}.  Section \ref{sec5} explains in more detail the connection of this walk to systems of territorial random walkers and section \ref{sec6} contains some concluding remarks.

\section{The model}
\label{sec2}

The starting point of our investigation consists of writing a master equation for the joint occupation probability $P_{n,m}(t)$ of the two particles being at site $n$ and $m$ at time $t$.  In relation to the territoriality problem, the two particles are the two boundaries that constitute a border, disregarding the presence of other territories.  Both cannot occupy the same site at the same time, but unless impeded by this constraint, at each hop the left-hand (right-hand) particle moves right (left) with probability $p$ and left (right) with probability $1-p$.  The hopping rate, i.e. hopping probability per unit time, is denoted by $F$ and the lattice spacing by $a$.  As particles may only hop to nearest-neighbour sites, we follow Aslangul's construction \cite{Aslangul} and write 

\begin{eqnarray}
{\rmd P_{n,m} \over \rmd t}(t) &= &2F[P_{n+1,m}(t)+P_{n,m-1}(t)](1-p)(1-\delta_{n,m-1})(1-\delta_{n,m}) \nonumber \\
                           &&+2F[P_{n-1,m}(t)+P_{n,m+1}(t)]p(1-\delta_{n,m+1})(1-\delta_{n,m}) \nonumber \\
                           &&-4F(1-\delta_{n,m+1})(1-\delta_{n,m})(1-p)P_{n,m}(t) \nonumber \\
                           &&-4F(1-\delta_{n,m-1})(1-\delta_{n,m})pP_{n,m}(t).
\label{master-eqn}
\end{eqnarray}

{\noindent}The term $1-\delta_{n,m}$, where $\delta$ is the Kronecker delta, represents the fact that two particles cannot hop from the same lattice site, whereas $1-\delta_{n,m\pm 1}$ represent the situations where both particles occupy adjacent lattice sites and so neither can move towards the other on the next hop. 

To seek the exact solution of (\ref{master-eqn}), it is convenient to use the generating function \cite{montroll1964} for $P_{n,m}(t)$, which is $f(\phi,\psi,t)=\sum_{n,m=-\infty}^{\infty}P_{n,m}(t)e^{in\phi}e^{im\psi}.$ 
The master equation (\ref{master-eqn}) implies the following relation for the generating function

\begin{eqnarray}
\fl{\rmd f(\phi,\psi,t) \over \rmd t}(t) = -4F\left\{f(\phi,\psi,t)-\int_0^{2\pi}\frac{\rmd\phi'}{2\pi}\left[1+\cos_{1-p}(\phi-\phi')\right]f(\phi',\phi+\psi-\phi',t)\right\} \nonumber \\
+2F(\cos_p\phi+\cos_{1-p}\psi)f(\phi,\psi,t)\nonumber \\
-2F\int_0^{2\pi}\frac{\rmd\phi'}{2\pi}f(\phi',\phi+\psi-\phi',t)\left[\cos_p\phi+\cos_{1-p}(\psi)\right]\nonumber \\
-2F\int_0^{2\pi}\frac{\rmd\phi'}{2\pi}f(\phi',\phi+\psi-\phi',t)\left[\cos_p(\phi')+\cos_{1-p}(\phi+\psi-\phi')\right],
\label{master-eqn-gf}
\end{eqnarray}

{\noindent}where we have introduced the notation $\cos_p(\theta)=p\rme^{\rmi\theta}+(1-p)\rme^{\rmi\theta}$ so that $\cos_{\frac{1}{2}}\equiv\cos$.

At time $t=0$ the particles occupy two lattice sites, denoted by $N_1$ and $N_2$.  Without loss of generality, assume $N_1<N_2$ and since the particles cannot cross, the particle starting at $N_1$ is referred to as the left-hand particle, the other is the right-hand particle.  By using this initial condition and setting $\theta=\phi+\psi$, the Laplace transform of (\ref{master-eqn-gf}) is

\begin{equation}
\tilde{f}(\phi,\theta-\phi,\epsilon)=g(\theta,\phi)+\sum_{i=1}^3 a_i(\theta,\phi) \int_0^{2\pi} \rmd\phi' b_i(\phi')\tilde{f}(\phi',\theta-\phi',\epsilon),
\label{master-eqn-gfl}
\end{equation}

{\noindent}where $\tilde{f}(\phi,\theta-\phi,\epsilon)= \int_0^\infty \rmd t f(\phi,\psi,t)\rme^{-\epsilon t}$ is the Laplace transform with variable $\epsilon$, and

\begin{eqnarray}
g(\theta,\phi) = {\rme^{\rmi(\theta-\phi)\Delta N}\rme^{\rmi N_1\theta} \over \epsilon + 2F[2-\cos_p\phi-\cos_{1-p}(\theta-\phi)]},  \nonumber \\
a_1(\theta,\phi) = {2F[2-\cos_p\phi-\cos_{1-p}(\theta-\phi)] \over \epsilon + 2F[2-\cos_p\phi-\cos_{1-p}(\theta-\phi)]},\qquad\quad b_1(\phi) = \frac{1}{2\pi}, \nonumber \\
a_2(\theta,\phi) = {2F(1-p)[2\rme^{\rmi\phi}-(1+\rme^{\rmi\theta})] \over \epsilon + 2F[2-\cos_p\phi-\cos_{1-p}(\theta-\phi)]},\quad\qquad b_2(\phi) = \frac{\rme^{-\rmi\phi}}{2\pi},  \nonumber \\
a_3(\theta,\phi) = {2Fp[2\rme^{-\rmi\phi}-(1+\rme^{-\rmi\theta})] \over  \epsilon + 2F[2-\cos_p\phi-\cos_{1-p}(\theta-\phi)]},\quad\qquad b_3(\phi) = \frac{\rme^{\rmi\phi}}{2\pi}, 
\end{eqnarray}

{\noindent}where $\Delta N=N_2-N_1$.  If, for a given value of $\theta$, we set $h(\phi)=\tilde f(\phi,\theta-\phi,\epsilon)$ and write $a_i(\phi)= a_i(\theta,\phi)$, $g(\phi)=g(\theta,\phi)$ to ease notation,  (\ref{master-eqn-gfl}) can be written in terms of the single variable $\phi$ as follows

\begin{equation}
h(\phi)=g(\phi)+\sum_{i=1}^3 a_i(\phi) \int_0^{2\pi} \rmd\phi' b_i(\phi')h(\phi').
\label{h-eqn}
\end{equation}

{\noindent}This is a Fredholm integral equation with degenerate kernel \cite{fredholmref}.  After some lengthy algebra (see Appendix A), the following solution is eventually found

\begin{eqnarray}
\fl\tilde{f}(\phi,\psi,\epsilon)=\rme^{\rmi N_1(\phi+\psi)}\Bigg\{\rme^{\rmi\Delta N\psi}\cos{\frac{\phi+\psi}{2}}(\rme^u-\rme^{(1-\Delta N)u}\rme^{\rmi{\Delta N}\frac{\phi-\psi}{2}}\bigg(\frac{p}{1-p}\bigg)^{\frac{{\Delta N}}{2}})\nonumber \\
+\rme^{(1-{\Delta N})u}\rme^{\rmi\psi}\rme^{\rmi({\Delta N}-1)\frac{\psi+\phi}{2}}\bigg(\frac{p}{1-p}\bigg)^{\frac{{\Delta N}}{2}}-\rme^{\rmi{\Delta N}\psi}\bigg(\frac{p}{1-p}\bigg)^{\frac{1}{2}}\Bigg\}\times \nonumber \\
\Bigg\{\bigg[\epsilon+2F(2-\cos_p\phi-\cos_{1-p}\psi)\bigg]\bigg[\rme^u \cos\frac{\phi+\psi}{2}-\bigg(\frac{p}{1-p}\bigg)^{\frac{1}{2}}\bigg]\Bigg\}^{-1}.
\label{complete-gen-laplace}
\end{eqnarray}

{\noindent}Here, $u$ is defined by the equation

\begin{equation}
\rme^u = \frac{Z+1+\sqrt{(Z+1)^2-4p(1-p)\cos^2\frac{\theta}{2}}}{2[p(1-p)]^{\frac{1}{2}}|\cos\frac{\theta}{2}|},
\label{etotheu}
\end{equation}

{\noindent}where $Z=\epsilon/4F$ and the branch of the square root function used here, and elsewhere throughout the text, is the one that takes real positive values when the argument is a positive real number.

\section{Asymptotic analysis}
\label{sec3}

In order to examine the asymptotics of the system, it is convenient to choose the initial conditions $N_1=0$, $N_2=1$, as this gives rise to a simpler form for (\ref{complete-gen-laplace})

\begin{equation}
\fl\tilde{f}(\phi,\psi,\epsilon)=\frac{\rme^{i\psi}|\cos\frac{\phi+\psi}{2}|}{4F[Z+1-\frac{1}{2}(\cos_p\phi+\cos_{1-p}\psi)]}.\frac{R(\phi+\psi,Z)\rme^{i\frac{\phi-\psi}{2}}-2(1-p)|\cos\frac{\phi+\psi}{2}|}{R(\phi+\psi,Z)-2(1-p)\cos^2\frac{\phi+\psi}{2}},
\label{complete-gen-laplaceN1}
\end{equation}

{\noindent}where $R(\theta,Z)=Z+1-\sqrt{(Z+1)^2-4p(1-p)\cos^2\frac{\theta}{2}}$.  This readily reduces to a result of Aslangul (equation 2.11 in \cite{Aslangul}) when $p=\frac{1}{2}$.

The marginal distribution for the left-hand (resp. right-hand) particle can be calculated by setting $\psi=0$ (resp. $\phi=0$).  For $-\pi<\phi<\pi$, we have the following expression for the generating function of the distribution of the left-hand particle in Laplace domain, when the right-hand particle can be anywhere else,

\begin{equation}
\tilde{f}(\phi,0,\epsilon)=\frac{\cos\frac{\phi}{2}}{4F(Z+\frac{1}{2}-\frac{1}{2}\cos_p\phi)}.\frac{R(\phi,Z)\rme^{\rmi\frac{\phi}{2}}-2(1-p)\cos\frac{\phi}{2}}{R(\phi,Z)-2(1-p)\cos^2\frac{\phi}{2}}.
\label{marginal-phi}
\end{equation}

{\noindent}This allows us to calculate the mean position $\langle x_1(\epsilon)\rangle$ of the left-hand particle in Laplace domain, by differentiating (\ref{marginal-phi}) with respect to $\phi$, multiplying by $-a\rmi$ and setting $\phi=0$, with the result

\begin{equation}
\fl\langle x_1(\epsilon)\rangle=\frac{a}{4\epsilon}\left(1-\frac{1}{\epsilon}\sqrt{\epsilon^2+8F\epsilon+16F^2(1-2p)^2}\right)+\frac{aF(2p-1)}{\epsilon^2}.
\label{MDcomplete}
\end{equation}

{\noindent}Differentiating (\ref{marginal-phi}) twice with respect to $\phi$, multiplying by $-a^2$ and again setting $\phi=0$ gives the second moment of the distribution

\begin{eqnarray}
\fl\langle x_1^2(\epsilon)\rangle=\frac{a^2}{4\epsilon}\left(1+\frac{8F}{\epsilon}-\frac{1}{\epsilon}\sqrt{\epsilon^2+8F\epsilon+16F^2(1-2p)^2}\right) \nonumber \\
+\frac{a^2(1-2p)}{\epsilon^3}\left(4F^2(1-2p)+F\sqrt{\epsilon^2+8F\epsilon+16F^2(1-2p)^2}\right).
\label{MSDcomplete}
\end{eqnarray}

{\noindent}By using the fact that $\mathcal{L}^{-1}[(\epsilon^2+2b\epsilon+b^2-a^2)^{-1/2}]=\rme^{-bt} I_0(at)$, where $\mathcal{L}^{-1}$ denotes the inverse Laplace transform and $I_\nu(z)$ a modified Bessel function of order $\nu$, expressions (\ref{MDcomplete}) and (\ref{MSDcomplete}) can be inverted exactly to give the respective formulae in time domain 

\begin{eqnarray}
\fl\langle x_1(\tau)\rangle=\frac{a}{4}\left(4(2p-2)\tau+8\sqrt{p(1-p)}\int_0^\tau {\rmd}s \frac{\tau-s}{s}\rme^{-4s}I_1[8\sqrt{p(1-p)}s]\right),
\label{MDexact} \\
\fl\langle x_1^2(\tau)\rangle=a^2\biggl((2-2p)\tau+2(1-2p)(2-2p)\tau^2 \nonumber \\
+2\sqrt{p(1-p)}\int_0^\tau {\rmd}s \frac{\tau-s-2(1-2p)(\tau-s)^2}{s}\rme^{-4s}I_1[8\sqrt{p(1-p)}s]\biggr),
\label{MSDexact}
\end{eqnarray}

{\noindent}where $\tau=tF$ is dimensionless time.  Denote by $x_1(\tau)$ and $x_2(\tau)$ the positions of the left- and right-hand particle respectively and let $d(\tau)=\langle x_2(\tau)-x_1(\tau)\rangle$ be the mean separation distance.  Since the second moments of the particles coincide and $\langle x_1(\tau)\rangle=-\langle x_2(\tau)\rangle$, it is convenient to denote by $\langle x^2(\tau) \rangle$ the second moment of either particle and by $\Delta x^2(\tau)=\langle x^2(\tau)-\langle x(\tau)\rangle^2\rangle$ the mean-square displacement.

If $p=\frac{1}{2}$ then the integrals in (\ref{MDexact}) and (\ref{MSDexact}) can be computed exactly \cite{Aslangul}.  For $p\neq \frac{1}{2}$, the integrals $\int_0^\infty \rmd s s^n \rme^{-4s} I_1[8\sqrt{p(1-p)}s]$ for $n=-1,0,1$ are the Laplace transforms of $t^n I_1[8\sqrt{p(1-p)}t]$ evaluated at the point where the Laplace variable is equal to $4$, that is

\begin{eqnarray}
{\mathcal L}\{t^{-1}I_1[8\sqrt{p(1-p)}t]\}(\epsilon)|_{\epsilon=4}=\frac{1-|1-2p|}{2\sqrt{p(1-p)}}, \nonumber \\
{\mathcal L}\{I_1[8\sqrt{p(1-p)}t]\}(\epsilon)|_{\epsilon=4}=\frac{1-|1-2p|}{8\sqrt{p(1-p)}|1-2p|}, \nonumber \\
{\mathcal L}\{tI_1[8\sqrt{p(1-p)}t]\}(\epsilon)|_{\epsilon=4}=\frac{\sqrt{p(1-p)}}{8|1-2p|^3}. \nonumber \\
\label{LaplaceTransformsofIBessel}
\end{eqnarray} 
 
{\noindent}Each of these three terms is finite for $p\neq \frac{1}{2}$, so this allows us to obtain asymptotic expressions for (\ref{MDexact}) and (\ref{MSDexact}) yielding the following expressions for $\tau\gg 1$:

\begin{equation}
\fl d(\tau)\approx\cases{\frac{p}{2p-1}a &\text{if $\frac{1}{2}<p<1$,}
\\
\sqrt{\frac{8}{\pi}}a\sqrt{\tau} & \text{if $p=\frac{1}{2}$,}
\\
4a(1-2p)\tau & \text{if $0<p<\frac{1}{2}$.}}
\label{MDlimit}
\end{equation}

\begin{equation}
\fl \langle x^2(\tau)\rangle\approx\cases{2a^2(1-p)\tau &\text{if $\frac{1}{2}<p<1$,}
\\
2a^2\tau & \text{if $p=\frac{1}{2}$,}
\\
4a^2(1-2p)^2\tau^2 & \text{if $0<p<\frac{1}{2}$.}}
\label{SecondMomentLimit}
\end{equation}

\begin{equation}
\fl \Delta x^2(\tau)\approx\cases{2a^2(1-p)\tau &\text{if $\frac{1}{2}<p<1$,}
\\
2a^2(1-\frac{1}{\pi})\tau & \text{if $p=\frac{1}{2}$,}
\\
2a^2 \tau & \text{if $0<p<\frac{1}{2}$.}}
\label{MSDlimit}
\end{equation}

\begin{figure}[htb]
\includegraphics[width=\columnwidth]{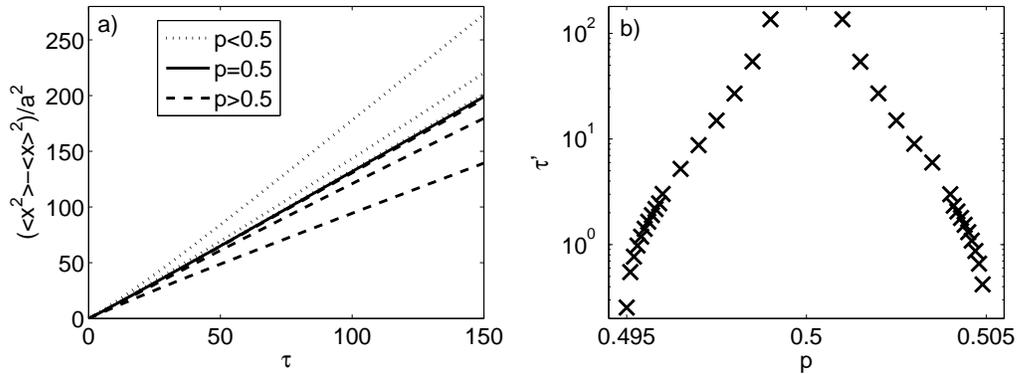}
\caption{Panel (a) shows the MSD as it varies through time for values of $p$ close to $\frac{1}{2}$, demonstrating when the MSD begins to split into three regimes, $p<\frac{1}{2}$, $p=\frac{1}{2}$, $p>\frac{1}{2}$.  Values of $p$ from the top curve to the bottom are $p=0.45$, $0.49$, $0.499$, $0.5$, $0.501$, $0.505$, $0.51$.  Panel (b) shows the timescale $\tau'$ beyond which the MSD curves for different values of $p$ diverge by more than $1\%$ from the curve for $p=\frac{1}{2}$.}
\label{ase_divergence}
\end{figure}

{\noindent}The different qualitative behaviours in both the MSD and the mean separation distance are now evident.  The limits $p\rightarrow \frac{1}{2}$ and $t\rightarrow \infty$ do not commute, so the asymptotic diffusion constant is very different in the case $p=\frac{1}{2}$ from the cases where $p$ is either just above or just below $\frac{1}{2}$.  Figure \ref{ase_divergence} shows the timescales in which the three regimes diverge from one another.

\begin{figure}[htb]
\includegraphics[width=\columnwidth]{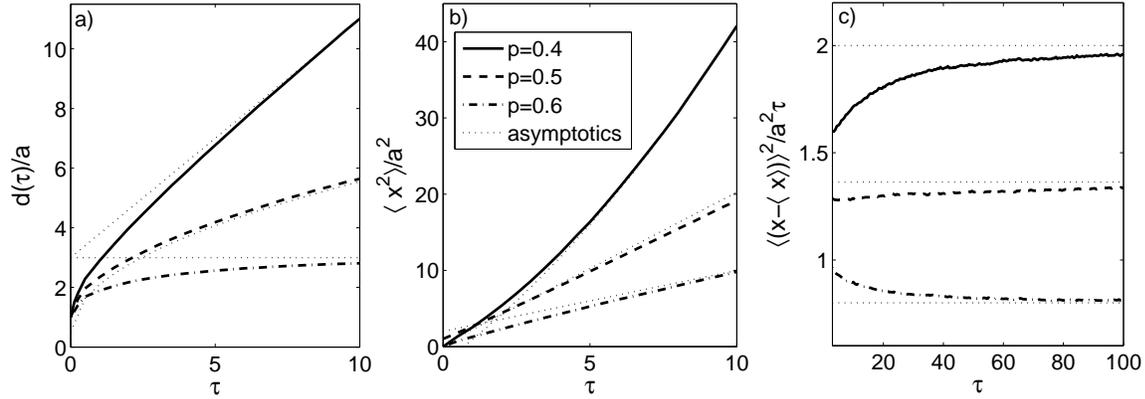}
\caption{Comparison of the asymptotic expressions from (\ref{MDlimit}), (\ref{SecondMomentLimit}) and (\ref{MSDlimit}) with average values of $10^6$ stochastic simulations of the system for $p=0.4,0.5,0.6$.  Panel (a) demonstrates how the mean distance between particles $d(\tau)$ exhibits qualitatively different behaviour in the three regions $p<\frac{1}{2}$, $p=\frac{1}{2}$ and $p>\frac{1}{2}$ when plotted against dimensionless time $\tau$.  Panel (b) shows the quadratic nature of the asymptotic second moment of a tagged particle when $p>\frac{1}{2}$, as compared with $p=\frac{1}{2}$ or $p<\frac{1}{2}$ when the second moments are asymptotically linear.  In panel (c), we see the particles reaching their asymptotic diffusion constants.}
\label{sim_vs_asymptotic}
\end{figure}

For $d(\tau)$, the different qualitative dependencies occur in the exponent of time so that for $p<\frac{1}{2}$ the displacement saturates, whereas for $\tau\geq\frac{1}{2}$ it increases.  Furthermore, this increase is linear for $p>\frac{1}{2}$ but sublinear when $p=\frac{1}{2}$.  Figure \ref{sim_vs_asymptotic} compares the various asymptotic expressions with simulation output for various $p$.

Conversely, at short times the behaviour of the system depends continuously on $p$.  For $\tau\ll 1$, considering only terms that are linear in $\tau$ we find:

\begin{eqnarray}
d(\tau)\approx  1+4a(1-p)\tau, \\
\langle x^2(\tau)\rangle \approx 2a^2(1-p)\tau .
\label{LimitShort}
\end{eqnarray}

\begin{figure}[htb]
\includegraphics[width=\columnwidth]{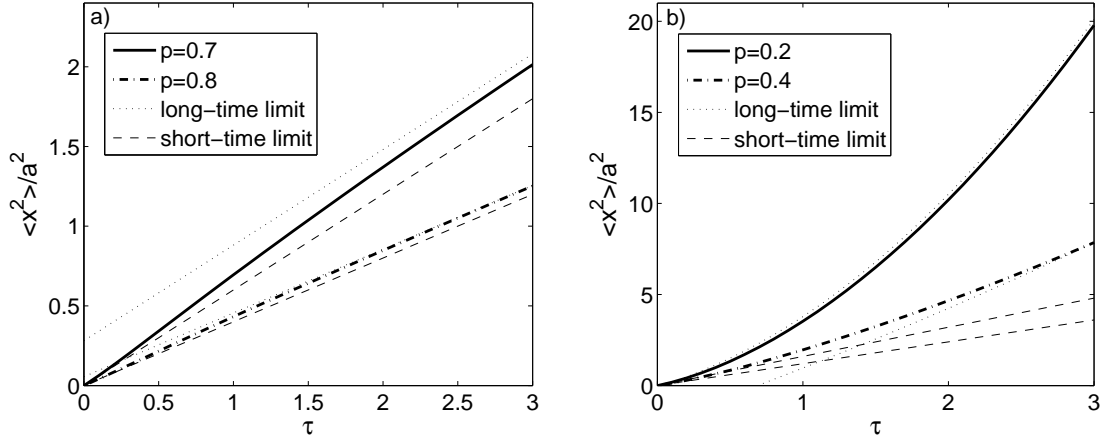}
\caption{Comparison of exact analytic expressions for the second moment (\ref{MSDexact}) with short-time (\ref{LimitShort}) and long-time (\ref{SecondMomentLimit}) approximations.  Panel (a) shows cases where $p>\frac{1}{2}$ and both approximate expressions are parallel.  As $p$ increases towards 1, the distance between the two approximations decreases and the curves converge faster towards the long-time expression.  Panel (b) shows cases where $p<\frac{1}{2}$.  The short-time approximations are linear whereas the long-time ones are quadratic.}
\label{analytic_vs_asymptotic}
\end{figure}

{\noindent}The second moment expression at short times differs from the corresponding long time expression by a constant for $p>\frac{1}{2}$ but by order $\tau$ for $p<\frac{1}{2}$.  Consequently, the shape of the second moment's evolution over time is very different for the two regions $p<\frac{1}{2}$ and $p>\frac{1}{2}$, despite their identical short-time approximations (see figure \ref{analytic_vs_asymptotic}).

\section{The continuum limit}
\label{sec4}

The transition to continuous space is made by taking the limits as $a \rightarrow 0$, $F\rightarrow \infty$, $N_1\rightarrow \infty$, $N_2\rightarrow \infty$ and $p\rightarrow \frac{1}{2}$ such that $D=a^2F$, $x_{1,0}=aN_1$, $x_{2,0}=aN_2$ and $v=2aF(2p-1)$.  Here, $D$ represents the diffusion constant, $x_{1,0}$ and $x_{2,0}$ the start positions of the left- and right-hand particles respectively and $v$ the velocity of one particle towards the other, the latter of which may be positive, zero or negative.  Also denote by $\Delta x_0 = x_{2,0}-x_{1,0}$ the distance between the two starting positions.

By setting $\phi=k_1a$ and $\psi=k_2a$, the aforementioned limit, is found for (\ref{complete-gen-laplace}) and denoted by $\tilde{\mathcal Q}(k_1,k_2,\epsilon)$:

\begin{eqnarray}
\fl \tilde{\mathcal Q}(k_1,k_2,\epsilon)=\frac{\rme^{\rmi{\Delta x_0}k_2}\rme^{\rmi x_{1,0}(k_1+k_2)}}{\epsilon + \rmi(k_2-k_1)v + \frac{D}{2}[(k_1+k_2)^2+(k_2-k_1)^2]} + \nonumber \\ \frac{\rmi\sqrt{\frac{D}{2}}(k_2-k_1)\rme^{\rmi{\Delta x_0}\frac{k_1+k_2}{2}}\rme^{\rmi x_{1,0}(k_1+k_2)}}{\epsilon+\rmi(k_2-k_1)v+\frac{D}{2}[(k_1+k_2)^2+(k_2-k_1)^2]}\times\nonumber \\ \frac{\exp\left[\frac{{\Delta x_0}}{\sqrt{2D}}\left(\frac{v}{\sqrt{2D}}-\sqrt{\epsilon+\frac{v^2}{2D}+\frac{D}{2}(k_1+k_2)^2}\right)\right]}{\sqrt{\epsilon+\frac{v^2}{2D}+\frac{D}{2}(k_1+k_2)^2}-\frac{v}{\sqrt{2D}}}.
\label{CtsbigNepsilonVelLim}
\end{eqnarray}

{\noindent}This reduces to a result of Aslangul (equation 3.1 in \cite{Aslangul}) by setting $v=0$, $x_{1,0}=0$ and $x_{2,0}=0$.  By using the identity 

\begin{equation}
\fl \mathcal{L}^{-1}\left[\frac{\rme^{-A(\sqrt{\epsilon+C}-B)}}{\sqrt{\epsilon+C}-B}\right]=\rme^{-Ct}\left\{\frac{\rme^{AB-\frac{A^2}{4t}}}{\sqrt{\pi t}}+B\left[1+{\mbox{erf}}\biggl(\frac{2Bt-A}{2\sqrt{t}}\biggr)\right]\rme^{B^2t}\right\},
\label{InvLaplaceGenFormula}
\end{equation}

{\noindent}from \cite{TheTable}, where ${\mbox{erf}}(z)$ is the error function, (\ref{CtsbigNepsilonVelLim}) can be Laplace inverted to give the following expression

\begin{eqnarray}
\fl {\mathcal Q}(k_1,k_2,t)=\rme^{-\rmi(k_2-k_1)vt-\frac{D}{2}[(k_1+k_2)^2+(k_2-k_1)^2]t}\Biggl\{\rme^{\rmi{\Delta x_0}k_2}+\sqrt{\frac{D}{2}}\rmi\rme^{\rmi{\Delta x_0}\frac{k_1+k_2}{2}}(k_2-k_1)\times\nonumber \\
\fl\int_0^t \rmd s\left[\frac{\rme^{-\frac{(\Delta x_0-2vs)^2}{8Ds}}}{\sqrt{\pi s}}+\frac{v}{\sqrt{2D}}{\mbox{erfc}}\biggl(\frac{{\Delta x_0}-2vs}{\sqrt{8Ds}}\biggr)\right]\rme^{-(\rmi(k_2-k_1)v-\frac{D}{2}(k_2-k_1)^2)s}\Biggr\}\rme^{\rmi x_{1,0}(k_1+k_2)},
\label{InvLaplaceCts}
\end{eqnarray}

{\noindent}where ${\mbox{erfc}}(z)$ is the complementary error function, ${\mbox{erfc}}(z)=1-{\mbox{erf}}(z)$.  In order to Fourier invert (\ref{InvLaplaceCts}) it is convenient to perform the double integral in the coordinates $K=k_1+k_2$ and $k=k_2-k_1$.  This procedure yields the joint probability distribution in continuous space and time 

\begin{eqnarray}
\fl Q(x_1,x_2,t)=\frac{\rme^{-\frac{(x_1-x_{1,0}-vt)^2}{4Dt}}}{\sqrt{4\pi Dt}}\frac{\rme^{-\frac{(x_2-x_{2,0}+vt)^2}{4Dt}}}{\sqrt{4\pi Dt}}+ \frac{\rme^{-\frac{(x_1-x_{1,0}+x_2-x_{2,0})^2}{8Dt}}}{\sqrt{8\pi D t}}\times\nonumber \\
\fl\int_0^t \rmd s \frac{[x_2-x_1+2v(t-s)]\rme^{-\frac{[x_2-x_1+2v(t-s)]^2}{8D(t-s)}}}{4D\sqrt{\pi (t-s)^3}}\left[\frac{\rme^{-\frac{(\Delta x_0-2vs)^2}{8Ds}}}{\sqrt{\pi s}}+\frac{v}{\sqrt{2D}}{\mbox{erfc}}\biggl(\frac{{\Delta x_0}-2vs}{\sqrt{8Ds}}\biggr)\right],
\label{InvLaplaceFourierCts}
\end{eqnarray}

{\noindent}where $Q(x_1,x_2,t)$ is the inverse Fourier transform of ${\mathcal Q}(k_1,k_2,t)$.  The first summand in (\ref{InvLaplaceFourierCts}) displays the short-time behaviour whereby the probability distribution of the left (right) particle can be approximated as a narrow Gaussian travelling right (left) at speed $v$ and the interaction between the two particles is minimal.  This interaction, represented by the second summand in (\ref{InvLaplaceFourierCts}), becomes more pronounced as time increases.

It turns out (Appendix B) that (\ref{InvLaplaceFourierCts}) is a solution to  the following Fokker-Planck equation that is obtained by taking the continuum limit of the discrete-space master equation (\ref{master-eqn-gf}) in the region $|n-m|>1$

\begin{equation}
\fl \qquad \frac{\partial Q}{\partial t}(x_1,x_2,t)=D\left(\frac{\partial^2}{\partial x_1^2}+\frac{\partial^2}{\partial x_2^2}\right)Q(x_{1},x_{2},t)+v\left(\frac{\partial}{\partial x_1}-\frac{\partial}{\partial x_2}\right)Q(x_{1},x_{2},t).
\label{diffusionEqn}
\end{equation}

{\noindent}However, this continuum limit is only valid for $x_1\neq x_2$.  Since the particles cannot cross, and therefore the probability density along $x_1=x_2$ must be zero, one can interpret this physically by imposing a zero-flux boundary condition along the line $x_1=x_2$ \cite{ambjornssonetal2008}, that is

\begin{equation}
\left[D\left(\frac{\partial}{\partial x_2}-\frac{\partial}{\partial x_1}\right)Q(x_1,x_2,t)+2vQ(x_1,x_2,t)\right]\Bigg|_{x_1=x_2}=0,
\label{ctsBdryCond}
\end{equation}

{\noindent}which is automatically satisfied by (\ref{InvLaplaceFourierCts}).  As such, the solution reduces to a result of Ambj\"ornsson {\it et al.} \cite{ambjornssonetal2008} in the case $v=0$, as well as Aslangul \cite{Aslangul} when additionally $x_{1,0}=0$ and $x_{2,0}=0$.  

To find expressions for the mean separation and MSD, an identical procedure to the discrete case is pursued (Appendix C), giving the following results

\begin{eqnarray}
\fl d(t)= {\Delta x_0}-vt\,{\mbox{erfc}}\left(\frac{2vt-{\Delta x_0}}{\sqrt{8Dt}}\right)-\frac{1}{\sqrt{\pi}}\int_0^t \rmd s\frac{2v^2s+{\Delta x_0}v-4D}{\sqrt{8Ds}}\rme^{-\frac{({\Delta x_0}-2vs)^2}{8Ds}},
\label{CtsbigNtimeMDVelLim}
\end{eqnarray}

\begin{eqnarray}
\fl \Delta x^2(t)= 2Dt-{\Delta x_0}vt+\frac{v^2t^2+{\Delta x_0}vt}{2}{\mbox{erfc}}\left(\frac{2vt-{\Delta x_0}}{\sqrt{8Dt}}\right)\nonumber \\
\fl+\frac{1}{\sqrt{\pi}}\int_0^t \rmd s \frac{v^2(2t-s)(2vs+{\Delta x_0})-8Dv(t-s)+{\Delta x_0}(2v^2s+{\Delta x_0}v-4D)}{4\sqrt{2Ds}}\rme^{-\frac{({\Delta x_0}-2vs)^2}{8Ds}}\nonumber \\
\fl -\left[\frac{vt}{2}{\mbox{erfc}}\left(\frac{2vt-{\Delta x_0}}{\sqrt{8Dt}}\right)+\frac{1}{\sqrt{\pi}}\int_0^t \rmd s\frac{2v^2s+{\Delta x_0}v-4D}{4\sqrt{2Ds}}\rme^{-\frac{({\Delta x_0}-2vs)^2}{8Ds}}\right]^2,
\label{CtsbigNtimeMSDVelLim}
\end{eqnarray}

{\noindent}where $\Delta x^2(t)$ is the MSD of either particle ($\Delta x_1^2(t)=\Delta x_2^2(t)$).  In the case $v=0$, the integrals in (\ref{CtsbigNtimeMDVelLim}) and (\ref{CtsbigNtimeMSDVelLim}) can be calculated exactly to give the following

\begin{eqnarray}
\fl d(t)=\sqrt{\frac{8D}{\pi}}\rme^{-\frac{{\Delta x_0}}{8Dt}}\sqrt{t}+{\Delta x_0}-{\Delta x_0}{\mbox{erfc}}\left({\Delta x_0}\sqrt{\frac{\pi}{2Dt}}\right), 
\label{CtsbigNtimeMDphalf}
\end{eqnarray}

\begin{eqnarray}
\fl \Delta x^2(t)=2Dt\left(1-\frac{1}{\pi}\rme^{-\frac{{\Delta x_0}}{8Dt}}\right) - {\Delta x_0}\sqrt{\frac{2Dt}{\pi}}\rme^{-\frac{{\Delta x_0}^2}{8Dt}}{\mbox{erf}}\left(\frac{{\Delta x_0}}{\sqrt{8Dt}}\right)+ \nonumber \\
\frac{{\Delta x_0}^2}{4}{\mbox{erfc}}\left(\frac{{\Delta x_0}}{\sqrt{8Dt}}\right)\left[2-{\mbox{erfc}}\left(\frac{{\Delta x_0}}{\sqrt{8Dt}}\right)\right].
\label{CtsbigNtimeMSDphalf}
\end{eqnarray}

{\noindent}For $v\neq 0$ on the other hand, the infinite integrals $\int_0^\infty \rmd s s^{n/2} \rme^{-\frac{({\Delta x_0}-4vs)^2}{8Ds}}$ for $n=-1,0,1$ are finite, so calculating them allows us to obtain asymptotic expressions for (\ref{CtsbigNtimeMDVelLim}) and (\ref{CtsbigNtimeMSDVelLim}) yielding the following expressions for $t\gg 1$:

\begin{equation}
\fl d(t)\approx\cases{\frac{D}{v} & \text{if $v>0$,}
\\
\sqrt{\frac{8Dt}{\pi}} & \text{if $v=0$,}
\\
-2vt &\text{if $v<0$.}}
\label{MDlimitCts}
\end{equation}

\begin{equation}
\fl \Delta x^2(t)\approx\cases{Dt & \text{if $v>0$,}
\\
2D(1-\frac{1}{\pi})t & \text{if $v=0$,}
\\
2Dt & \text{if $v<0$.}}
\label{MSDlimitCts}
\end{equation}

{\noindent}This contrasts with the small-time limit $t\ll1$, whereby $d(t)\approx {\Delta x_0}-2vt$ and $\Delta x^2(t) \approx 2Dt$ for any $v$.  

Notice that the $v>0$ ($v=0$, $v<0$) cases of (\ref{MDlimitCts}) and (\ref{MSDlimitCts}) are simply the continuous-space limits of the $p>\frac{1}{2}$ ($p=\frac{1}{2}$, $p<\frac{1}{2}$) cases in the discrete-space expressions (\ref{MDlimit}) and (\ref{MSDlimit}).  For example, in the case $p>\frac{1}{2}$ from (\ref{MDlimit}), by setting $a^2\tau=Dt$ and $v=2aF(2p-1)$, we obtain $d(t)=2Dp/v$ and by taking the limit $p\rightarrow \frac{1}{2}$ one recovers the continuous asymptotic result $d(t)\approx D/v$ reported in (\ref{MDlimitCts}).  Likewise, setting $a^2\tau=Dt$ in the case $p>\frac{1}{2}$ from (\ref{MSDlimit}) and by taking the limit $p\rightarrow \frac{1}{2}$ one recovers the continuous asymptotic result $\Delta x^2(t)\approx Dt$ from (\ref{MSDlimitCts}).  

\section{Connection to territorial random walkers}
\label{sec5}

In \cite{GPH2}, simulation analysis of the many-bodied, non-Markovian system of territorial random walkers demonstrated that the asymptotic generalised (because of single-file phenomena) diffusion constant of a territory border depends on an interplay between the so-called {\it active scent time} $T_{\rm AS}$, the time for which a scent mark is recognised by conspecifics as an active territory cue, and the animal population density $\rho$.  Specifically, the border diffusion constant decays exponentially as the dimensionless parameter $Z=T_{\rm AS}R\rho^2 a^2$ is increased, where $R$ is the rate of the animal's movement between lattice sites, separated by distance $a$.  Part of the purpose of the present study is to gain a deeper insight into why this phenomenon is observed.

\begin{figure}[htb]
\includegraphics[width=\columnwidth]{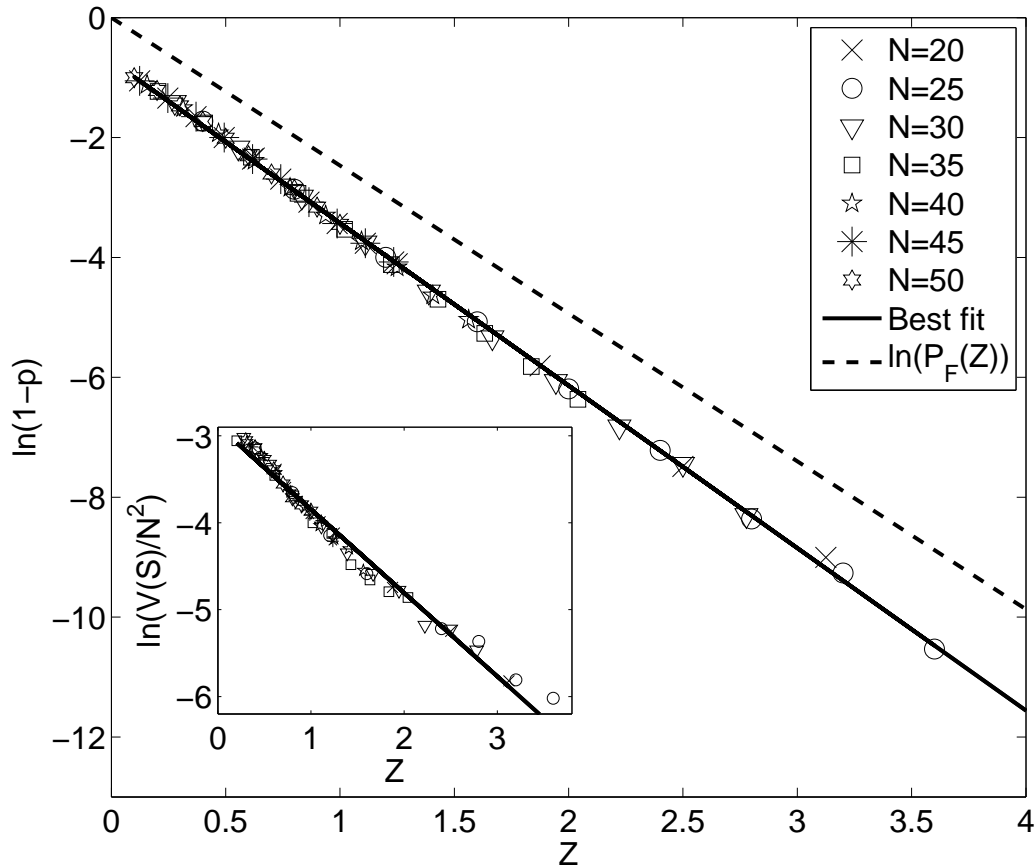}
\caption{The relationship between the value of $p$ measured from simulations of a system of 1D territorial random walkers and the dimensionless quantity $Z$ defined in secton \ref{sec5}.  This is compared with the probability $P_F(Z)$ that the animal fails to traverse a territory of average width $1/\rho$ within a time $T_{\rm AS}$.  In order to measure $p$ from the simulations, the number of times a boundary moved towards the adjacent boundary were counted, and divided by the total number of times that the boundary moved.  The equations for the curves are $1-p=0.49\rme^{-2.7Z}$ and $P_F(Z)=\rme^{-\pi^2Z/4}$, where $\pi^2/4\approx 2.5$.  The inset shows how the variance $V(S)$ of the territory size $S$ decays as $Z$ increases, used in the main text to explain the discrepancy between the rate of the exponential decays of the two curves in the main plot.}
\label{ase_sim_p}
\end{figure}

In the territorial random walk system, $p$ is the probability that, if there is a gap between two adjacent boundaries, that gap will decrease in length the next time a boundary moves.  Such a probability is clearly always greater or equal to $\frac{1}{2}$ on average, otherwise the territories would fail to maintain a positive average width.  For such values of $p$, equation (\ref{MSDlimit}) shows that the asymptotic diffusion constant of a boundary is proportional to $1-p$.  

When we measure the value of $1-p$ directly from the simulations, we see that it also decays exponentially as $Z$ increases, suggesting that calculating $p$ is of fundamental importance in understanding why the border diffusion constant decays exponentially as $Z$ increases.  This relationship between $p$ and $Z$ can be explained as follows.  First we observe that the probability of a boundary decaying is likely to be closely related to the probability of an animal traversing its territory within a time $T_{\rm AS}$.  To this end, we calculate the first-passage probability ${\mathcal F}(t)$ for an animal to traverse a territory of average length $1/\rho$, which corresponds in our lattice system to an integer $N=1/a\rho$ sites.  Since this is equivalent to the situation where the animal starts at a reflecting boundary has to traverse to the other, absorbing boundary, the asymptotic value of this first-passage probability is calculated in \cite{redner} to be ${\mathcal F}(t)\sim\rme^{-\pi^2Rt/4 N^2}$.  Therefore the probability $P_{\rm F}(Z) \propto \int_{T_{\rm AS}}^\infty \rmd t{\mathcal F}(t)$ of failing to traverse the territory within a time $T_{\rm AS}$ is approximately $\rme^{-\pi^2Z/4}$.  In figure \ref{ase_sim_p}, $P_{\rm F}(Z)$ is plotted alongside the simulation measurements for $1-p$ showing that both decay exponentially with increasing $Z$ and with similar exponents.  


To explain the small discrepancy in the two exponents, we make the observation that as $Z$ is increased, the variance in the territory width decreases in an approximately exponential fashion (inset figure 4).  Because of the $N^2$ dependence of the mean first passage time to cross the territory \cite{redner}, the mean first passage time increases as the variance in the territory width increases.  Therefore for a fixed $\rho$, the actual mean first passage time to traverse a territory decreases as $Z$ increases, whereas above we have assumed that the first passage probability is always equal to that of a territory of average width.  This has the effect of causing the probability $1-p$ to decrease with $Z$ slightly faster than in the analytic estimation.  In other words the curve of $P_{\rm F}(Z)$ decays slightly slower than the curve of simulation measurements of $1-p$.



\section{Conclusions}
\label{sec6}

The propogator for a system of two anti-symmetric, biased random walkers on an infinite lattice is computed exactly.  We characterize the bias via the parameter $0<p<1$, representing the probability for the walkers of moving away from each other, towards each other or with no bias at all.  These three distinct physical scenarios depend, respectively, on the value of $p$ being less than, greater than or equal to $\frac{1}{2}$.  When $p$ is less than $\frac{1}{2}$, the walkers drift away from one another with a mean displacement that is asymptotically linear in time.  At $p=\frac{1}{2}$ the random walkers still drift apart, although the mean displacement scales as the square root of time.  For $p>\frac{1}{2}$, the distance between the particles saturates.  The asymptotic saturation distance comes about because of two opposing tendencies in the walkers: a drift towards one another, given by the amount of bias in the walkers' movement, and the magnitude of their intrinsic diffusion. 

The corresponding propagator for the continuum limit is also computed exactly by setting $D=a^2F$ and $v=2aF(2p-1)$ in the limits $a \rightarrow 0$, $F\rightarrow \infty$ and $p\rightarrow \frac{1}{2}$.  It is the solution of a Fokker-Planck equation with zero-flux boundary conditions whenever the particles meet.

A motivation for this study is part of a programme to understand systems of territorial random walkers \cite{GPH1}.  A first step in that direction has been the study \cite{GPH2} of a reduced one-body dynamics for the movement of a single animal within subdiffusing territorial borders attached by springs.  The present work represents an additional step in the formulation of a simplified model of the non-Markovian dynamics of territorial random walkers, whereby the fine scale dynamics of a border are studied through the analysis of its constituent adjacent boundaries modelled as anti-symmetric random walkers with a bias probability $p>1/2$. Future work will involve studying the effects of having a sequence of interlinked, randomly moving borders, as sketched in figure \ref{territory_explanation}, with the left and right boundary of each territory connected by springs.

\section*{Acknowledgements}

This work was partially supported by the EPSRC grant number EP/E501214/1 (JRP and LG) and by the Dulverton Trust (SH).  We thank three anonymous referees for their useful comments.

\appendix
\section*{Appendix A}
\label{appA}
\setcounter{section}{1}

Since (\ref{h-eqn}) from the main text is a Fredholm equation with degenerate kernel \cite{fredholmref}, its solution is a linear combination of the quantities $\gamma_i=\int_0^{2\pi}\rmd \phi'b_i(\phi')h(\phi')$ for $i=1,2,3$, which satisfy the following system of equations

\begin{eqnarray}
\gamma_1(1-\alpha_{11})-\gamma_2\alpha_{12}-\gamma_3\alpha_{13}&=&\beta_1,\nonumber \\
-\gamma_1\alpha_{21}+\gamma_2(1-\alpha_{22})-\gamma_3\alpha_{23}&=&\beta_2,\nonumber \\
-\gamma_1\alpha_{31}-\gamma_2\alpha_{32}+\gamma_3(1+\alpha_{33})&=&\beta_3.
\label{system-sym-eqs}
\end{eqnarray}

{\noindent}The various $\beta_i$'s and $\alpha_{ij}$'s can be calculated as $\beta_i=\int_0^{2\pi}\rmd \phi'b_i(\phi')g(\phi')$ and $\alpha_{ij}=\int_0^{2\pi}\rmd \phi'b_i(\phi')a_j(\phi')$ to yield the following expressions

\begin{eqnarray}
\beta_1&=\frac{\rme^{\rmi N_1\theta}\rme^{\frac{\rmi{\Delta N}\theta}{2}}\rme^{-{\Delta N}u}}{8F(1-p)\cos\frac{\theta}{2}\sinh u}\left(\frac{p}{1-p}\right)^{\frac{{\Delta N}-1}{2}},\nonumber \\
\beta_2&=\frac{\rme^{\rmi N_1\theta}\rme^{\frac{\rmi({\Delta N}-1)\theta}{2}}\rme^{-({\Delta N}+1)u}}{8F(1-p)\cos\frac{\theta}{2}\sinh u}\left(\frac{p}{1-p}\right)^{\frac{{\Delta N}}{2}},\nonumber \\
\beta_3&=\frac{\rme^{\rmi N_1\theta}\rme^{\frac{\rmi({\Delta N}+1)\theta}{2}}\rme^{-({\Delta N}-1)u}}{8F(1-p)\cos\frac{\theta}{2}\sinh u}\left(\frac{p}{1-p}\right)^{\frac{{\Delta N}-2}{2}},\nonumber \\
\alpha_{11}&=\frac{[p(1-p)]^{-\frac{1}{2}}-2\rme^{-u}\cos\frac{\theta}{2}}{2\cos\frac{\theta}{2}\sinh u},\nonumber \\
\alpha_{21}&=\frac{\rme^{-\frac{\rmi\theta}{2}}\rme^{-u}\{(1-p)^{-1}-2[p/(1-p)]^{\frac{1}{2}}\cos\frac{\theta}{2}\cosh u\}}{2\cos\frac{\theta}{2}\sinh u},\nonumber \\
\alpha_{31}&=\frac{\rme^{\frac{\rmi\theta}{2}}\rme^{-u}\{p^{-1}-2[(1-p)/p]^{\frac{1}{2}}\cos\frac{\theta}{2}\cosh u\}}{2\cos\frac{\theta}{2}\sinh u},\nonumber \\
\alpha_{12}&=\frac{\rme^{\frac{\rmi\theta}{2}}\{\rme^{-u}[(1-p)/p]-[(1-p)/p]^{\frac{1}{2}}\cos\frac{\theta}{2}\}}{2\cos\frac{\theta}{2}\sinh u},\nonumber \\
\alpha_{22}&=\frac{[(1-p)/p]^{\frac{1}{2}}-\rme^{-u}\cos(\frac{\theta}{2})}{2\cos\frac{\theta}{2}\sinh u},\nonumber \\
\alpha_{32}&=\frac{[(1-p)/p]\rme^{\rmi\theta}\rme^{-u}\{[(1-p)/p]^{\frac{1}{2}}\rme^{-u}-\cos\frac{\theta}{2}\}}{2\cos\frac{\theta}{2}\sinh u},\nonumber \\
\alpha_{13}&=\frac{\rme^{-\frac{\rmi\theta}{2}}\{\rme^{-u}[p/(1-p)]-[p/(1-p)]^{\frac{1}{2}}\cos\frac{\theta}{2}\}}{2\cos\frac{\theta}{2}\sinh u},\nonumber \\
\alpha_{23}&=\frac{[p/(1-p)]\rme^{-\rmi\theta}\rme^{-u}\{\rme^{-u}[p/(1-p)]^{\frac{1}{2}}-\cos\frac{\theta}{2}\}}{2\cos\frac{\theta}{2}\sinh u},\nonumber \\
\alpha_{33}&=\frac{[p/(1-p)]^{\frac{1}{2}}-\rme^{-u}\cos(\frac{\theta}{2})}{2\cos\frac{\theta}{2}\sinh u}.
\end{eqnarray}

{\noindent}In these equations $u$ is defined by (\ref{etotheu}) in the main text.  Solving the system of equations (\ref{system-sym-eqs}) eventually gives 

\begin{eqnarray}
\gamma_1=\gamma_2=0\nonumber \\
\gamma_3=\frac{\rme^{\rmi N_1\theta}\rme^{\rmi\frac{({\Delta N}+1)\theta}{2}}\rme^{-{\Delta N}u}[p/(1-p)]^{\frac{{\Delta N}-2}{2}}}{4F\{(1-p)\cos\frac{\theta}{2}-[p(1-p)]^{\frac{1}{2}}\rme^{-u}\}}
\label{gammas}
\end{eqnarray}

{\noindent}Plugging these values for $\gamma_i=\int_0^{2\pi}\rmd\phi'b_i(\phi')h(\phi')$ into (\ref{h-eqn}) in the main text gives the expression for the generating function of the system's probability distribution in Laplace domain.

\appendix
\section*{Appendix B}
\label{appB}
\setcounter{section}{2}

In order to solve (\ref{diffusionEqn}) with boundary condition (\ref{ctsBdryCond}) from the main text, it is convenient to convert to coordinates $x_s=x_2-x_1$ and $x_c=(x_1+x_2)/2$ so that $x_s$ is the separation distance between the particles and $x_c$ is the centroid.  This allows us to write (\ref{diffusionEqn}) as 

\begin{equation}
\frac{\partial R}{\partial t}(x_c,x_s,t)=D\left(\frac{1}{2}\frac{\partial^2}{\partial x_c^2}+2\frac{\partial^2}{\partial x_s^2}\right)R(x_c,x_s,t)+2v\frac{\partial R}{\partial x_s}(x_{c},x_{s},t),
\label{diffusionEqnCS}
\end{equation}

{\noindent}where $R(x_c,x_s,t)=Q(x_1,x_2,t)$.  The flux vector of equation (\ref{diffusionEqnCS}) is 

\begin{eqnarray}
J=-\left[2D\frac{\partial R}{\partial x_s}(x_c,x_s,t)+2vR(x_c,x_s,t)\right]
\label{BoundaryCondsR}
\end{eqnarray}

{\noindent}so the zero-flux boundary condition mentioned in the main text is $\hat{n}\cdot J|_{x_s=0}=0$ where $\hat{n}$ is a unit normal to the line $x_s=0$ \cite{ambjornssonetal2008}. By writing $R(x_c,x_s,t)=R_c(x_c)R_s(x_s)$, (\ref{diffusionEqnCS}) becomes 

\begin{eqnarray}
\frac{\partial R_c}{\partial t}(x_c,t)=\frac{D}{2}\frac{\partial^2 R_c}{\partial x_c^2}(x_c,t),
\label{diffusionEqnRC}
\end{eqnarray}

{\noindent}and

\begin{eqnarray}
\frac{\partial R_s}{\partial t}(x_s,t)=2D\frac{\partial^2 R_s}{\partial x_s^2}(x_s,t)+2v\frac{\partial R_s}{\partial x_s}(x_s,t),
\label{diffusionEqnRS}
\end{eqnarray}

{\noindent}with the boundary condition

\begin{eqnarray}
\left[D\frac{\partial R_s}{\partial x_s}(x_s,t)+vR_s(x_s,t)\right]\Bigg|_{x_s=0}=0.
\label{BoundaryCondsRs}
\end{eqnarray}

{\noindent}The solution to (\ref{diffusionEqnRC}) is a Gaussian and the solution to (\ref{diffusionEqnRS}) with boundary condition (\ref{BoundaryCondsRs}) can be found in e.g. \cite{polyanin} with the result

\begin{eqnarray}
R_s(x_c,t)=\frac{\rme^{-\frac{(x_c-x_{c,0})^2}{2Dt}}}{\sqrt{2\pi D t}},
\label{diffusionEqn2Solns1}
\end{eqnarray}
\begin{eqnarray}
R_c(x_s,t)=H(x_s)\Bigg[&\frac{\rme^{-\frac{(x_s-x_{s,0}+2vt)^2}{8Dt}}}{\sqrt{8\pi D t}} + \frac{\rme^{\frac{v}{2D}(x_{s,0}-vt-x_s)}\rme^{-\frac{(x_s+x_{s,0})^2}{8Dt}}}{\sqrt{8\pi D t}}\nonumber \\
&+\frac{v}{2D}{\mbox{erfc}}\left(\frac{x_s+x_{s,0}-2vt}{\sqrt{8Dt}}\right)\rme^{-\frac{vx}{D}}\Bigg],
\label{diffusionEqn2Solns2}
\end{eqnarray}

{\noindent}where $x_{s,0}=\Delta x_0$ and $x_{c,0}=(x_{1,0}+x_{2,0})/2$ are the initial conditions, and $H(x)$ is the Heaviside step function ($H(x)=0$ if $x<0$, $H(x)=1$ if $x\geq 0$).  The solution to (\ref{diffusionEqn}) from the main text can now be written down as

\begin{eqnarray}
\fl Q(x_1,x_2,t)=H(x_2-x_1)\frac{\rme^{-\frac{(x_1-x_{1,0}+x_2-x_{2,0})^2}{2Dt}}}{\sqrt{8\pi D t}}\Bigg[\frac{\rme^{-\frac{(x_2-x_{2,0}-x_1+x_{1,0}+2vt)^2}{8Dt}}}{\sqrt{2\pi D t}} +\nonumber \\
\frac{\rme^{\frac{v}{2D}(x_{2,0}-x_{1,0}-vt-x_2+x_1)}\rme^{-\frac{(x_2-x_1+x_{2,0}-x_{1,0})^2}{8Dt}}}{\sqrt{2\pi D t}}
+\nonumber \\ 
\frac{v}{D}{\mbox{erfc}}\left(\frac{x_2-x_1+x_{2,0}-x_{1,0}-2vt}{\sqrt{8Dt}}\right)\rme^{-\frac{vx}{D}}\Bigg]
\label{diffusionEqnSoln}
\end{eqnarray}

{\noindent}In order to show that (\ref{diffusionEqnSoln}) is equivalent to (\ref{InvLaplaceFourierCts}) from the main text, the following integral is calculated

\begin{eqnarray}
\fl I(x_s,t)=\int_0^t \rmd s \frac{[x_s+2v(t-s)]\rme^{-\frac{[x_s+2v(t-s)]^2}{8D(t-s)}}}{4D\sqrt{\pi (t-s)^3}}\Bigg[&\frac{\rme^{-\frac{(\Delta x_0-2vs)^2}{8Ds}}}{\sqrt{\pi s}}+\nonumber \\
&\frac{v}{\sqrt{2D}}{\mbox{erfc}}\biggl(\frac{{\Delta x_0}-2vs}{\sqrt{8Ds}}\biggr)\Bigg].
\label{FullCtsExprIntegral}
\end{eqnarray}

{\noindent}Since this is the sum of two convolutions in time, its Laplace transform can be found by using the identity ${\mathcal L}[f*g]={\mathcal L}[f]{\mathcal L}[g]$, where the asterix denotes the convolution $f*g=\int_0^t \rmd sf(s)g(t-s)$.  Since  

\begin{eqnarray}
\frac{[x_s+2v(t-s)]\rme^{-\frac{[x_s+2v(t-s)]^2}{8D(t-s)}}}{4D\sqrt{\pi (t-s)^3}} = -\frac{\partial}{\partial x_s}\frac{\rme^{-\frac{[x_s+2v(t-s)]^2}{8D(t-s)}}}{\sqrt{\pi (t-s)}},\nonumber
\label{DiffWRTxs}
\end{eqnarray}

{\noindent}the Laplace transform of $I(x_s,t)$ can be written as

\begin{eqnarray}
\fl {\mathcal L}[I(x_s,t)]=-\frac{\partial}{\partial x_s}\Bigg[&\frac{\rme^{-\frac{v(x_s-\Delta x_0)}{2D}}\rme^{\frac{|x_s|+\Delta x_0}{\sqrt{2D}}\sqrt{\epsilon +\frac{v^2}{2D}}}}{\epsilon +\frac{v^2}{2D}}+\nonumber \\ 
&\frac{v}{\sqrt{2D}}\frac{\rme^{-\frac{v(|x_s|+\Delta x_0)}{2D}}\rme^{\frac{|x_s|+
\Delta x_0}{\sqrt{2D}}\left(\sqrt{\epsilon+\frac{v^2}{2}}-\frac{v}{\sqrt{2D}}\right)}}{\Big(\epsilon +\frac{v^2}{2}\Big) \left(\sqrt{\epsilon+\frac{v^2}{2}}-\frac{v}{\sqrt{2D}}\right)}\Bigg].
\label{ILaplace}
\end{eqnarray}

{\noindent}By repeatedly using the formula (\ref{InvLaplaceGenFormula}) from the main text, expression (\ref{ILaplace}) can be Laplace inverted to give

\begin{eqnarray}
\fl I(x_s,t)=-\frac{\partial}{\partial x_s}\Biggl[&\rme^{-\frac{v}{2D}(|x_s|+x_s)}{\mbox{erfc}}\left(\frac{|x_s|+\Delta x_0-2vt}{\sqrt{8Dt}}\right)\Biggr].
\label{ILaplaceInverted}
\end{eqnarray}

{\noindent}Performing the differentiation with respect to $x_s$ gives

\begin{eqnarray}
I(x_s,t)=&\frac{v}{2D}\left({\mbox{sgn}}(x_s)+1\right)\rme^{-\frac{v}{2D}(|x_s|+x_s)}{\mbox{erfc}}\left(\frac{|x_s|+\Delta x_0 -2vt}{\sqrt{8Dt}}\right)+\nonumber \\
&{\mbox{sgn}}(x_s)\frac{\rme^{\frac{v}{2D}(\Delta x_0-x_s-vt)}\rme^{\frac{(|x_s|+\Delta x_{0})^2}{8Dt}}}{\sqrt{2\pi D t}}
\label{IxtIntegrated}
\end{eqnarray}

{\noindent}where ${\mbox{sgn}}(x)$ is the sign of $x$ (${\mbox{sgn}}(x)=-1$ if $x<0$ and ${\mbox{sgn}}(x)=1$ if $x\geq 0$).  After replacing the second term of (\ref{InvLaplaceFourierCts}) from the main text with $I(x_s,t)$ one can show that the continuum limit of (\ref{complete-gen-laplace}) is indeed the solution of the Fokker-Planck equation (\ref{diffusionEqn}) with the above mentioned zero-flux boundary conditions.

\appendix
\section*{Appendix C}
\label{appC}
\setcounter{section}{3}

Since the values of $d(t)$ and $\Delta x^2(t)$ depend only on the initial condition $\Delta x_0$ and not the specific values of $x_{1,0}$ and $x_{2,0}$, calculations are simplified by assuming $x_{1,0}=0$.  The marginal probability distribution for the left-hand (right-hand) particle in Fourier-Laplace domain is found by setting $k_2=0$ ($k_1=0$) in equation (\ref{CtsbigNepsilonVelLim}).  Focussing on the left-hand particle gives the following expression

\begin{eqnarray}
\fl \tilde{\mathcal Q_1}(k_1,\epsilon)=\frac{1}{\epsilon - \rmi k_1v + Dk_1^2} -\frac{\rmi\rme^{\rmi{\Delta x_0}\frac{k_1}{2}}k_1}{\epsilon-\rmi k_1v+\frac{D}{2}k_1^2} \frac{\sqrt{\frac{D}{2}}\rme^{\frac{{\Delta x_0}}{\sqrt{2D}}\left(\frac{v}{\sqrt{2D}}-\sqrt{\epsilon+\frac{v^2}{2D}+\frac{D}{2}k_1^2}\right)}}{\sqrt{\epsilon+\frac{v^2}{2D}+\frac{D}{2}k_1^2}-\frac{v}{\sqrt{2D}}}.
\label{CtsbigNepsilonVelLimMarg}
\end{eqnarray}

{\noindent}{\noindent}This allows us to calculate the mean position $\langle x_1(\epsilon)\rangle$ of the left-hand particle in Laplace domain, by differentiating (\ref{marginal-phi}) with respect to $k_1$, multiplying by $-\rmi$ and setting $k_1=0$

\begin{equation}
\langle x_1(\epsilon)\rangle=\frac{v}{\epsilon^2}-\frac{\sqrt{D}\rme^{\frac{{\Delta x_0}}{\sqrt{2D}}\left(\frac{v}{\sqrt{2D}}-\sqrt{\epsilon+\frac{v^2}{2D}}\right)}}{\epsilon\sqrt{2}\left(\sqrt{\epsilon+\frac{v^2}{2D}}-\frac{v}{\sqrt{2D}}\right)}.
\label{MDcompleteCts}
\end{equation}

{\noindent}Differentiating (\ref{CtsbigNepsilonVelLimMarg}) twice with respect to $k_1$, multiplying by $-1$ and again setting $k_1=0$ gives the second moment of the distribution

\begin{eqnarray}
\langle x_1^2(\epsilon)\rangle=\frac{2v^2}{\epsilon^3}+\frac{2D}{\epsilon^2}-\frac{({\Delta x_0}\epsilon+2v)\sqrt{D}\rme^{\frac{{\Delta x_0}}{\sqrt{2D}}\left(\frac{v}{\sqrt{2D}}-\sqrt{\epsilon+\frac{v^2}{2D}}\right)}}{\epsilon^2\sqrt{2}\left(\sqrt{\epsilon+\frac{v^2}{2D}}-\frac{v}{\sqrt{2D}}\right)}.
\label{MSDcompleteCts}
\end{eqnarray}

{\noindent}By using the formula (\ref{InvLaplaceGenFormula}) from the main text, (\ref{MDcompleteCts}) and (\ref{MSDcompleteCts}) can be inverted exactly to give the respective formulae in time domain.  Performing the same calculations for the right-hand particle allows us to find the following expressions for the mean separation and MSD

\begin{eqnarray}
\fl d(t)={\Delta x_0}-v\int_0^t\rmd s{\mbox{erfc}}\left(\frac{2vs-{\Delta x_0}}{\sqrt{8Ds}}\right)+\sqrt{\frac{2D}{\pi}}\int_0^t \rmd s\frac{\rme^{-\frac{({\Delta x_0}-2vs)^2}{8Ds}}}{\sqrt{s}},
\label{MDBeforeIBP}
\end{eqnarray}

\begin{eqnarray}
\fl \Delta x^2(t)= 2Dt-{\Delta x_0}vt+v^2\int_0^t \rmd s(t-s){\mbox{erfc}}\left(\frac{2vs-{\Delta x_0}}{\sqrt{8Ds}}\right)-\nonumber \\
\sqrt{\frac{D}{2\pi}}\int_0^t \rmd s \frac{{\Delta x_0}+2v(t-s)}{\sqrt{s}}\rme^{\frac{({\Delta x_0}-2vs)^2}{8Ds}}- \nonumber \\
\left[\frac{v}{2}\int_0^t \rmd s {\mbox{erfc}}\left(\frac{2vs-{\Delta x_0}}{\sqrt{8sD}}\right)-\sqrt{\frac{D}{2\pi}}\int_0^t \rmd s\frac{\rme^{-\frac{({\Delta x_0}-2vs)^2}{8Ds}}}{\sqrt{s}}\right]^2.
\label{MSDBeforeIBP}
\end{eqnarray}

{\noindent}Expressions (\ref{CtsbigNtimeMDVelLim}) and (\ref{CtsbigNtimeMSDVelLim}) from the main text are obtained by applying the formula $\int_0^t \rmd s {\mbox{erf}}\left(\frac{As-B}{\sqrt{s}}\right)=t\,{\mbox{erf}}\left(\frac{As-B}{\sqrt{s}}\right)-\frac{1}{\sqrt{\pi}}\int_0^t \rmd s\frac{As+B}{\sqrt{s}}\rme^{-\left(\frac{As-B}{\sqrt{s}}\right)^2}$ throughout (\ref{MDBeforeIBP}) and (\ref{MSDBeforeIBP}).

\end{document}